\documentclass[preprint]{elsarticle}
\usepackage{lineno,xcolor,rotating,setspace,amsthm,amsmath,amsfonts}
\graphicspath{{figures/}}
\usepackage[ruled]{algorithm2e}
\SetKwComment{Comment}{$\triangleright$\ }{}
\DontPrintSemicolon
\usepackage{hyperref}
\hypersetup{colorlinks=true, linkcolor=blue, filecolor=magenta, urlcolor=blue, citecolor=blue, pdfauthor=Zachary P. Neal}
\modulolinenumbers[1]
\newcommand{\doi}[1]{\href{http://dx.doi.org/#1}{https://doi.org/#1}}
\newcommand{\DOIprefix}{}
\biboptions{authoryear}  % APA style

\begin{document}

\begin{frontmatter}
\title{The duality of networks and groups: Models to generate two-mode networks from one-mode networks}

\author{Zachary P. Neal\corref{cor1}}
\ead{zpneal@msu.edu}
\address{Michigan State University}

\begin{abstract}
Focus theory describes how shared memberships, social statuses, beliefs, and places can facilitate the formation of social ties, while two-mode projections provide a method for transforming two-mode data on individuals' memberships in groups into a one-mode network of their possible social ties. In this paper, I explore the opposite process: how social ties can facilitate the formation of groups, and how a two-mode network can be generated from a one-mode network. Drawing on theories of team formation, club joining, and organization recruitment, I propose three models that describe how such groups might emerge from the relationships in a social network. I show that these models can be used to generate two-mode networks that have characteristics commonly observed in empirical two-mode social networks, and that they encode features of the one-mode networks from which they were generated. I conclude by discussing these models' limitations, and future directions for theory and methods concerning group formation.
\end{abstract}

\end{frontmatter}

%\linenumbers

%\noindent\rev{Note to reviewers: This is a submission of a previously rejected \textit{Social Networks} submission that the editors welcomed as a new submission. Although this is a new submission, my responses to the original reviewers are included a supplementary file.}

\section{Introduction}
A natural question in the social networks literature has been: \textit{Where do social networks come from?} The answers have been diverse, and contributions have taken the form of both theoretical propositions for underlying mechanisms such as homophily \citep[e.g.,][]{mcpherson2001birds} and statistical frameworks for testing these propositions \citep[e.g.,][]{robins2007introduction, snijders2010introduction}. Focus theory proposed that social networks come from groups such as parties or clubs that present opportunities for individuals to meet and form ties by focusing social activity \citep{feld1981focused}. However, this raises the obvious question: \textit{Where do such groups come from?}

There is a duality of social networks and groups, such that networks can come from groups, but groups can also come from networks. A sketch of this duality was already present in the initial articulation of focus theory \citep{feld1981focused}. However most subsequent work has examined how networks emerge from groups, while neglecting how groups emerge from networks. In this paper, I aim to elaborate the second half of this duality. Drawing from a range of disciplinary contexts, I develop three models for how groups can emerge from social networks: as teams \citep{guimera2005team}, as clubs \citep{backstrom2006group,schaefer2022youth}, and as organizations \citep{mcpherson2004blau}. While these models offer insight into how groups can emerge from networks, they also contribute to the methodological literature as two-mode network generative models, which currently ``are practically non-existent'' in the literature \citep[][p. 3]{vasques2020latent}.

The remainder of the paper is organized in five sections. In section \ref{sec:background} I briefly review the theories and methods available for understanding how networks and groups co-evolve. Then, in section \ref{sec:models} I introduce three models for how groups can emerge from social networks. In section \ref{sec:generative} I use simulations to show that these models can be used to generate two-mode networks that have characteristics commonly observed in empirical two-mode social networks, and illustrate how the generated two-mode networks encode features of the one-mode network from which they were generated. Finally, in section \ref{sec:discussion} I conclude by considering these models' limitations and their potential applications for building and testing both theories and methods.

\section{Background}
\label{sec:background}
\cite{wellman1988structural} warned that ``the world is composed of networks, not groups'' (p. 37). This claim may have gone too far, but did highlight that networks and groups are different. A group is a collection of individuals who might be socially cohesive or not; its internal social structure is unspecified. In contrast, a network is a social structure among individuals who might cluster into discrete sets or not; its members' categorical affiliations are unspecified. Accordingly, it is possible to study just networks, just groups, or how they influence each other. Table \ref{tab:processes} shows how the formation of networks and groups has been studied, and highlights where the present study fits among these lines of research.

\begin{table}
\begin{tabular}{lp{8cm}}
\hline
Process & Representative theories and methods \\
\hline
Network $\to$ Network & Balance, Homophily, Preferential attachment, Reciprocity, Transitivity \citep{fuhse2022networks,yap2015does}; Temporal exponential random graph models \citep[TERGM;][]{krivitsky2014separable} and Stochastic actor oriented models \citep[SOAM;][]{snijders2010introduction} \\
Groups $\to$ Groups & Diffusion of innovation theory \citep{rogers1995diffusion}; Two-mode/Bipartite ERGM \citep[BERGM;][]{wang2013exponential} and SOAM \citep{schaefer2022youth} \\
Groups $\to$ Network & Focus theory \citep{feld1981focused}; Two-mode projections \citep{breiger1974duality} and backbones \citep{neal2014backbone} \\
Network $\to$ Groups & Mentioned but not developed in focus theory \citep{feld1981focused}; ``generative models are practically non-existent'' \citep[p. 3][]{vasques2020latent} \\
\hline
\end{tabular}
\caption{The evolution and formation of networks and groups}
\label{tab:processes}
\end{table}

Within the field of (social) network analysis, perhaps the most widely studied formative process involves the evolution of networks. Many mechanisms have been hypothesized to explain how network ties form, and how ties that are present or absent at time 1 impact the presence or absence of ties at time 2 \citep{fuhse2022networks,yap2015does}. For example, ties may form and networks may evolve through a process of preferential attachment, such that new ties tend to be formed with already well-connected others. Ties may also form through processes of balance that promote friendship cycles (i.e., A $\to$ B $\to$ C $\to$ A), or of status seeking that prohibit them. Among the most intuitive tie formation processes are those that unfold when individuals share something in common. Ties can form through homophily when individuals share an interest or demographic characteristic, through propinquity when they share a space, or through transitivity when they share a set of common friends. While theories of tie formation are well developed, so too are formal statistical methods for modeling the evolution of networks, which include temporal exponential random graph models \citep[TERGM;][]{krivitsky2014separable} and stochastic actor oriented models \citep[SOAM;][]{snijders2010introduction}.

The evolution of groups has also been well-studied. Groups can evolve in several ways, including expanding by merging with other groups or when new members join, and shrinking by splintering into smaller groups or when existing members leave. Because the process of groups at time 1 evolving into groups at time 2 does not explicitly implicate networks, research on group evolution often does not draw on network theories or methods. One notable exception might be diffusion of innovation theory \citep{rogers1995diffusion}, which seeks to explain how and when members of the non-adopter group become members of the adopter group, which can depend in part on networks \citep[e.g.][]{valente1996social}. Because group memberships can be represented using two-mode or bipartite networks, extensions to statistical methods for modeling network evolution have proven useful for modeling group evolution \citep{wang2013exponential,schaefer2022youth}.

Much attention has been devoted to how networks or groups evolve over time, but it is also possible to study how one emerges from the other. Focus theory \citep{feld1981focused} hypothesized that networks emerge from individuals' shared membership in groups or other `foci', which are ``social, psychological, legal, or physical entity around which joint activities are organized'' \citep[][p. 1016]{feld1981focused}. The interactions that take place in these groups ``bring people together in a mutually rewarding situation'' because they are focused on something that is shared, and therefore these interactions are ``positively valued'' \citep[][p. 1017]{feld1981focused}. Through these positively valued interactions, the participants ``develop positive sentiments toward each other'' and thus positive affective ties \citep[][p. 1026]{feld1981focused}. \cite{feld1981focused} summarized the process by explaining that ``As a consequence of interaction associated with their joint activities, individuals whose activities are organized around the same focus will tend to become interpersonally tied and form a cluster'' \citep[][p. 1016]{feld1981focused}. Two-mode projections \citep{breiger1974duality} and backbones \citep{neal2014backbone} represent generative models that formalize this hypothesis and explicitly show how networks can emerge from groups by transforming a two-mode network into a one-mode network

Although focus theory is traditionally viewed as explaining how networks emerge from groups, \cite{feld1981focused} also acknowledged that groups can emerge from networks, noting that ``Once there is a tie between two individuals, these individuals will tend to find and develop new foci around which to organize their joint activity.'' Indeed, his diagram of the dynamics of the focus model is cyclical, with groups creating ties, which in turn create new groups. \cite{schaefer2022youth} recently provided empirical evidence of this process, finding that direct influence from friends was the single most important exogeneous predictor of whether a high school student would form or join a new extracurricular activity. However, none of focus theory's twenty propositions deal with how or when networks emerge from groups, and corresponding ``generative models are practically not-existent'' \citep[][p. 3]{vasques2020latent}. It is these gaps that the present work seeks to fill, thereby filling in the underdeveloped second half of focus theory, and developing the complement to two-mode projection.

\section{Groups from networks}
\label{sec:models}
Focus theory \citep{feld1981focused} and bipartite projection \citep{breiger1974duality,neal2014backbone} already offer a detailed description of how networks might emerge from groups. In this section, I propose three models for how a group might emerge from a network. Each model represents a simplified implementation of a theory about the formation of a specific type of group: teams, clubs, and organizations. For each model, I first present the motivating theory, then describe the model, and provide a concrete illustration of a group forming from a network according to the model. Pseudocode algorithm representations of each model are presented as appendices.

\subsection{Teams Model}
The \textit{teams} model derives from an existing model of team formation. \cite{guimera2005team} suggested that the individuals who form teams in a given setting are embedded in a ``complex network [that is] the medium in which future collaborations will develop'' (p. 697-8). That is, teams emerge from an existing social network. In their original model, all teams had a fixed size $m$. Each of the $m$ positions on a newly forming team were filled based on probabilities $p$ and $q$. Specifically, a position was filled with: (a) a new person joining the setting from an unlimited pool of outsiders with probability $1-p$, (b) a person who is already a member of the setting with probability $p(1-q)$, or (c) a person who is already a member of the setting and is connected in the social network to individuals on the new team with probability $pq$. Their model was dynamic because outsiders join the setting over time, and because each new team contributes to the network that influences the formation of future teams. It is also complex because it is parameterized by three values $m$, $p$, and $q$.

The teams model is a modification of \citeauthor{guimera2005team}'s (\citeyear{guimera2005team}) complex dynamic model, and allows teams of varying size to emerge from a static network based on a single parameter $p$. Given an existing social network, cliques represent sets of colleagues who all know or interact with one another, and who therefore might form a team. Each new team emerges from one of these cliques, but can involve changes in membership. Because some of the clique's members (i.e. incumbents) may be unavailable or lack the necessary skills for the newly-forming team's task, they must be replaced by others (i.e. newcomers). The model outcome depends on a parameter that specifies the probability with which incumbents are retained ($p$), rather than replaced by newcomers on new teams ($1-p$). Accordingly, the parameter $p$ controls how closely the memberships of new teams will match the memberships of cliques in an existing social network. When $p = 1$, where incumbents are always retained, the teams model reduces to the model described by \cite{guillaume2004bipartite}, where teams are equivalent to cliques. A pseudocode algorithm of the teams model is provided in the \textit{appendix}.

Figure \ref{fig:team} provides a concrete example. Suppose the network on the left is a network among colleagues in an academic department, and the clique \{A,B,C\} represents a set of colleagues who know each other, perhaps because they worked together on a grant proposal. A new three-member team is emerging from this group to submit a new proposal. Because they are the ones initiating the new team's formation, the first position on the new team must be filled by one of the group's incumbents \{A,B,C\}. In this example, the first position is filled by incumbent A. The remaining two positions on the team are filled by selecting incumbents with probability $p$, and selecting newcomers with probability $1-p$. In this example, the second position is filled by newcomer D, while the third position is filled by incumbent B. The new team \{A,D,B\} could be the outcome of a situation in which newcomer D replaced incumbent C to address reviewers' concerns with the earlier proposal.

\begin{figure}
    \centering
    \includegraphics[width=\textwidth]{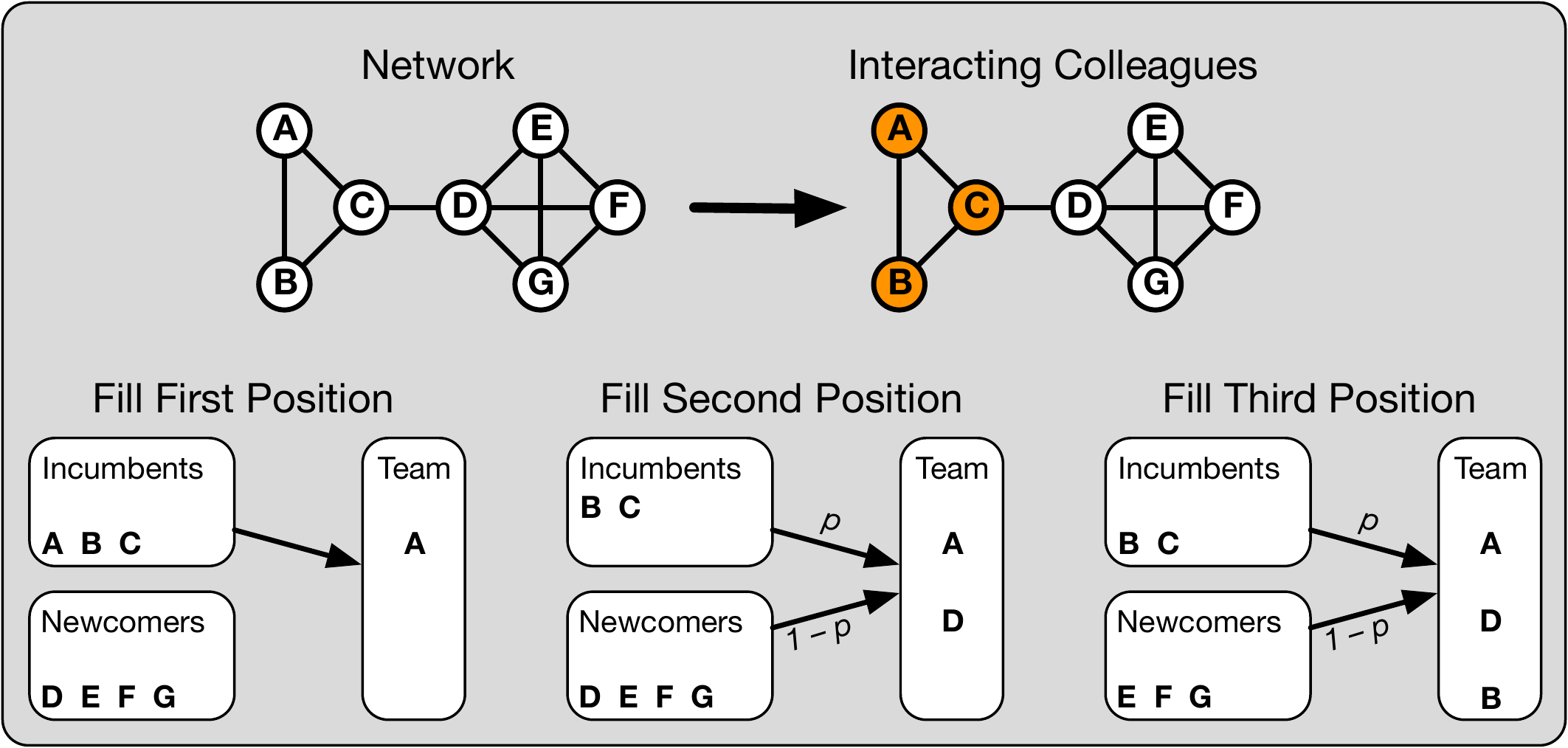}
    \caption{Example of group formation via the teams model.}
    \label{fig:team}
\end{figure}

\subsection{Clubs Model}
The \textit{clubs} model is informed by findings about how social groups such as clubs form in both online and offline social networks. \cite{backstrom2006group} examined 19 characteristics of the group and potential joiner in two online social networks (LiveJournal and DBLP), while \cite{schaefer2022youth} considered 8 mechanisms that drive high school students to join extracurricular activities. Both studies found that the probability of joining a group depends on the number of friends one already has in the group. Additionally, \cite{backstrom2006group} found that the probability of joining a group also depends on the proportion of friends in the group who are friends with each other. 

While these studies focused on individuals joining existing groups, their findings also have implications for the density of a newly forming group. The fact that $i$ tends to join a group when she already has many friends $j$ in the group increases the group's density by increasing the likelihood of $i$--$j$ edges. Additionally, the fact that $i$ tend to join a group when her friends $j$ in the group are friends with each other increases the group's density by increasing the likelihood of $j$--$j$ edges. Therefore, groups whose initial formation is guided by the conditions identified by \cite{backstrom2006group} and \cite{schaefer2022youth} will be cohesive and have a relatively higher density than the overall network. From this implication, the clubs model views clubs as forming via an agglomeration process: a clique serves as the seed of a potential club. Then, seeking to establish a viable club, members recruit their friends, who join on the condition that the club would maintain a minimum density $p$. Accordingly, $p$ functions as a parameter that controls the club formation process. When $p = 1$, where new members join only if the new club would be a clique, the clubs model reduces to the model described by \cite{guillaume2004bipartite}, where groups are equivalent to cliques. A pseudocode algorithm of the clubs model is provided in the \textit{appendix}.

Figure \ref{fig:club} provides a concrete example, where $p = 0.7$. Suppose the network on the left is a friendship network, within which a group of friends \{D,E,F,G\} (a randomly selected clique) wishes to start a book club. To make their book club viable, they must recruit other friends to participate. The challenge is that these friends are socially anxious and only feel comfortable in group settings where at least 70\% of the members are friends with each another. Initially C is the only candidate because they are friends with existing book club members. The book club attempts to recruit C, and C decides to join because doing do would result in a book club in which 70\% of the members are friends with each other. Once C is a member, A and B become candidates for recruitment. The book club attempts to recruit A first, however A declines to join because doing so would yield a book club in which only 53\% of members are friends with each other. The book club's attempt to recruit B is unsuccessful for the same reason. Thus, the new book club's members are \{D,E,F,G,C\}.

\begin{figure}
    \centering
    \includegraphics[width=\textwidth]{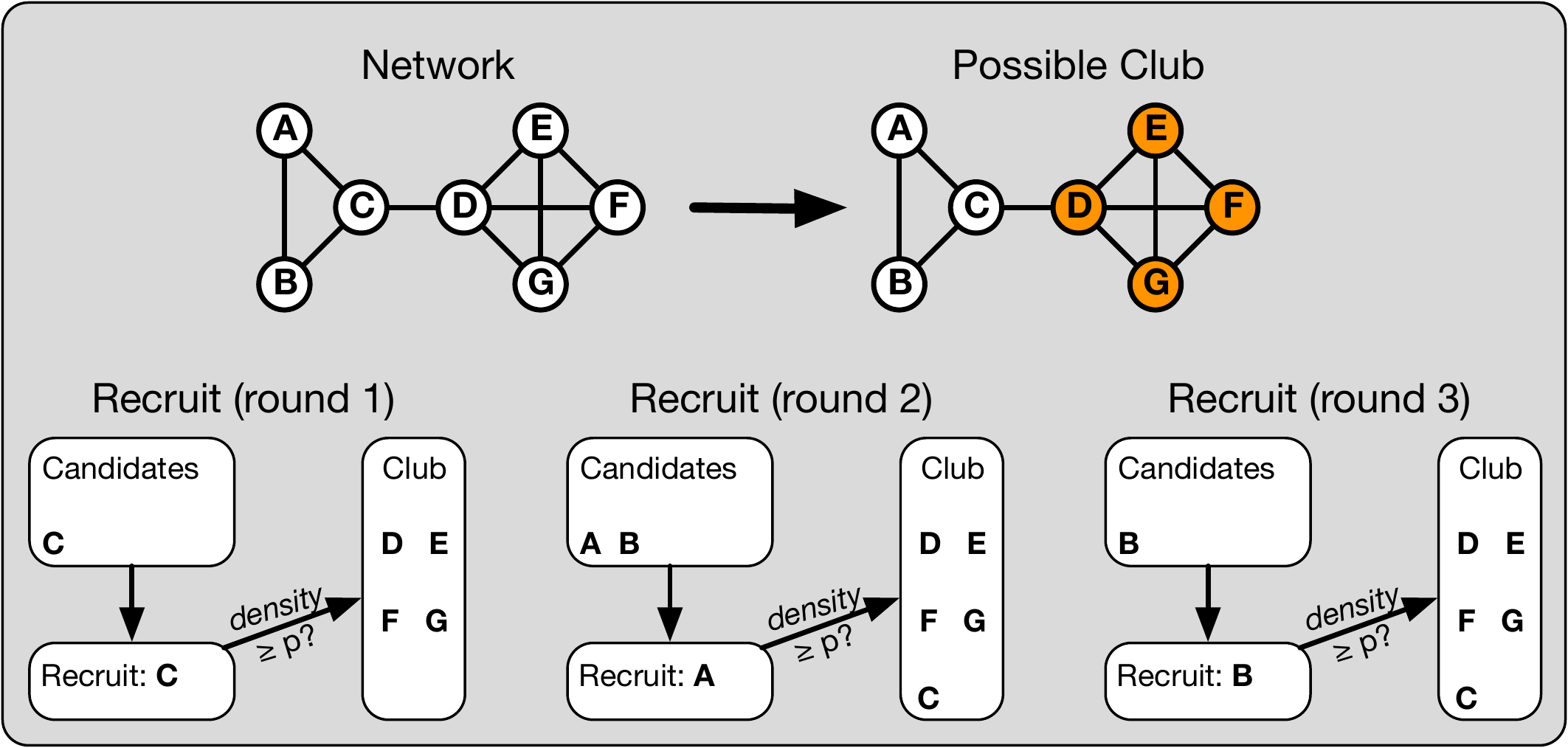}
    \caption{Example of group formation via the clubs model.}
    \label{fig:club}
\end{figure}

\subsection{Organizations Model}
The \textit{organizations} model mirrors the Blau space model of organizational recruitment \citep{mcpherson1983ecology,mcpherson2004blau}. Blau space is a multidimensional space within which individuals are located based on their sociodemographic characteristics. As \cite{mcpherson2004blau} explains, Blau space has two important properties: it ``at once organizes the social interactions among individuals, and structures the opportunities for the formation of social entities that are associated with individuals in that space'' (p. 267). First, it organizes social interactions because individuals who are sociodemographically similar are located nearby in the space and, according to the principle of homophily \citep{mcpherson2001birds}, are therefore more likely to interact with each other. This implies that network ties will tend to be local within Blau space. Second, it structures the formation of social entities because organizations recruit members from specific regions in this space, known as niches \citep{popielarz2007niche}. For example, a youth yachting league might recruit its members from the region located at the lower end of the age dimension, but the upper end of the family wealth dimension.

The organizations model does not attempt to formalize all aspects of niche or organizational ecology theories \citep{popielarz2007niche,shi2017member}, but instead is a simplification that incorporates only two central elements: individuals' positions in an unobserved Blau space derived from their distances in a social network, and organizations' recruitment of members from niches in this space. Individuals' locations in Blau space can be estimated by embedding network geodesic distances in a $d$-dimensional space \citep{freeman1983spheres,peli2006networks}. While $d$ can take any value between $1$ and $N-1$, where $N$ is the number of nodes in the network, I use a two-dimensional space because social networks tend to have low dimensionality \citep{freeman1983spheres,bonato2014dimensionality}, because many dimensions of social distinction are highly correlated (e.g., income and education), and because Blau space analysis is typically performed on low dimensionality spaces \citep{genkin2018blaunet}. Organizations have $d$-dimensional circular niches within this space that reflect the type of member they seek to recruit \citep{peli2006networks,suh2017negligible}. Organizations niche sizes vary, however most organizations are narrow-niche specialists, while a few are wide-niche generalists \citep{carroll1985concentration}. An organization's success at recruiting members depends on whether the prospective members are inside its niche (with probability $p$) or outside its niche \citep[with probability $1-p$; ][]{popielarz1995edge}. Accordingly, $p$ serves as a parameter that controls the importance of niche location in individuals' joining behavior. A pseudocode algorithm of the organizations model is provided in the \textit{appendix}.

Figure \ref{fig:organization} provides a concrete example. Suppose the network on the left is a social network of neighborhood friends. The geodesic distances between individuals in the network can be used to embed them in a 2-dimensional space via multidimensional scaling. Friends (e.g., C \& D) are close together in this space, while friends-of-friends (e.g., A \& D) are further apart in this space, and friends-of-friends-of-friends (e.g., A \& F) are furthest apart. The sociodemographic characteristics described by these two dimensions are unknown, but perhaps they are income and education; notice the two dimensions are highly correlated. A multi-level marketing company selling beauty products aims to recruit sales associates; its niche is people who have less income and education, which includes four people. It recruits each person inside this niche with probability $p$, and in this example successfully recruits A, B, and D. Because its niche included four people, it aims to still recruit a fourth sales associate. It attempts to recruit those nearest the niche first, with probability $1-p$. In this example, it fails to recruit E, but successfully recruits G, at which point recruitment ends. This yields a neighborhood sales team of \{A,B,D,G\}.

\begin{figure}
    \centering
    \includegraphics[width=\textwidth]{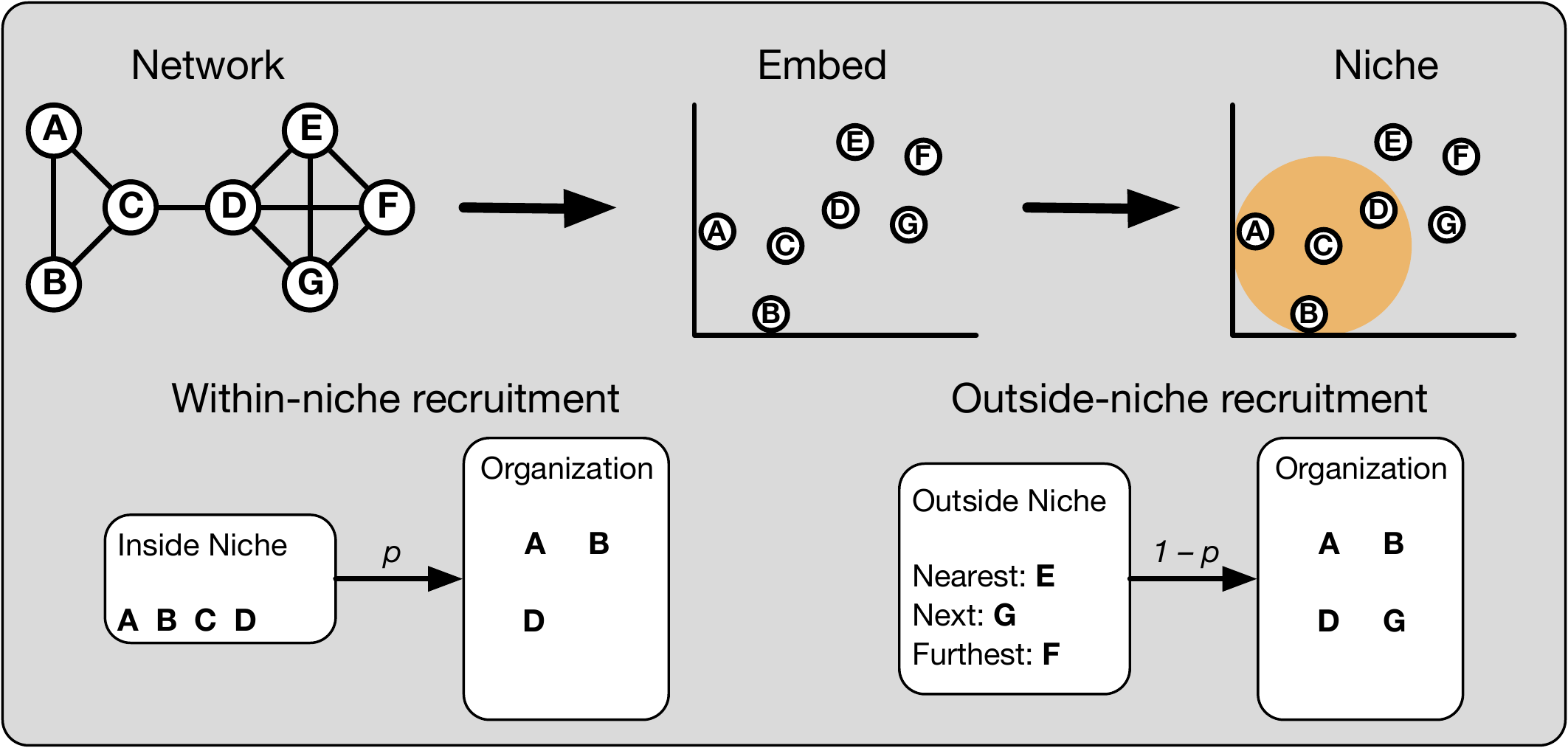}
    \caption{Example of group formation via the organizations model.}
    \label{fig:organization}
\end{figure}

\section{Two-mode generative models}
\label{sec:generative}
The models introduced in section \ref{sec:models} each describe how one new group might emerge from an existing social network. However, if they are applied repeatedly on the same social network, they can also be viewed as generative models because they can generate two-mode networks representing group memberships from one-mode networks representing social networks. Many one-mode network generative models already exist, including the Erd\H{o}s-R\'enyi model for generating random graphs \citep{erdosrenyi}, the Watts-Strogatz model for generating small world graphs \citep{watts1998collective}, and the Barab\'asi-Albert model for generating scale free graphs \citep{barabasi1999emergence}. However, as \cite{vasques2020latent} observe, ``when it comes to bipartite networks [i.e., two-mode networks] -- a class of network frequently encountered in social systems, among others -- generative models are practically non-existent'' (p. 3).

Generative models are not designed to simulate actual processes in the world, but instead are designed to reproduce observed empirical patterns using simple mechanisms. For example, the Watts-Strogatz model simply involves randomly re-wiring edges in a regular lattice. While this does not simulate how social networks actually form (people don't randomly swap friends), it does generate networks with characteristics observed in empirical social networks (e.g. clustering). Likewise, as two-mode generative models, these are not designed to simulate actual group formation processes, which are likely quite complex. Instead, they are designed to generate two-mode networks that have characteristics observed in empirical two-mode social networks, and that encode features of the one-mode networks from which they were generated. In this section, I explore the extent to which they achieve these goals. The generative models are implemented in the \texttt{incidence.from.adjacency()} function in the \texttt{incidentally} package for \textsf{R} \citep{incidentally}. The code necessary to reproduce the results reported in this section is available at 
\url{https://osf.io/eyua4/}.
%\url{https://osf.io/eyua4/?view_only=7b731b732cb540efb05947a535672e9f}.

\subsection{Reproducing empirical patterns}
\label{sec:experiments}
One way to evaluate these generative models involves examining whether they generate two-mode networks that have characteristics commonly observed in empirical two-mode social networks. Although much attention has been devoted to identifying the typical or universal properties of social networks \citep[e.g., clustering, degree distributions][]{watts1998collective,barabasi1999emergence}, relatively little work has examined the typical or universal properties of two-mode social networks. However, three characteristics are commonly observed: positively skewed agent degree distributions, positively skewed group degree distributions, and short cycles.

In a two-mode network generated by these models, the agent degree distribution captures the number of groups with which each agent is associated. Across many empirical contexts, this degree distribution tends to be positively skewed because most people are associated with just a few groups, while some people are associated with many groups. For example, most students participate in just a few extracurricular activities while some participate in many \citep{schaefer2022youth}, most legislators sponsor just a few bills while some sponsor many \citep{neal2020sign}, most women attend just a few parties while some attend many \citep{davis1941deep}, and most authors write just a few papers while some write many \citep{filho2020transitivity}.

The group degree distribution captures the number of agents associated with each group. Again, across many empirical contexts this degree distribution tends to be positively skewed because most groups have just a few members, while some groups have many members. For example, most extracurricular activities have just a few participants while some have many \citep{schaefer2022youth}, most bills are sponsored by just a few legislators while some are sponsored by many \citep{neal2020sign}, most parties have just a few attendees while some have many \citep{davis1941deep}, and most papers have just a few authors while some have many \citep{filho2020transitivity}.

Finally, empirical two-mode networks typically contain more four-cycles than would be expected at random. A four-cycle occurs when two nodes of one type are both connected to the same two nodes of another type, or in this context, two people are both members of the same two groups. \cite{filho2020transitivity} demonstrated this pattern in three author-paper networks and one member-board network, arguing that it helps explain the strong ties observed in social networks due to shared groups. Drawing on this empirical pattern, \cite{schaefer2022youth} explicitly hypothesized observing the formation of four-cycles through a mechanism they called ``co-member influence,'' whereby high school students join the same new extracurricular activities as co-members of their existing extracurricular activities. Indeed, this is such an important property of two-mode networks that \cite{saracco2015randomizing} explicitly sought to count and control four-cycles (calling them X-motifs) in their bipartite null models.

\begin{figure}
    \centering
    \includegraphics[width=\textwidth]{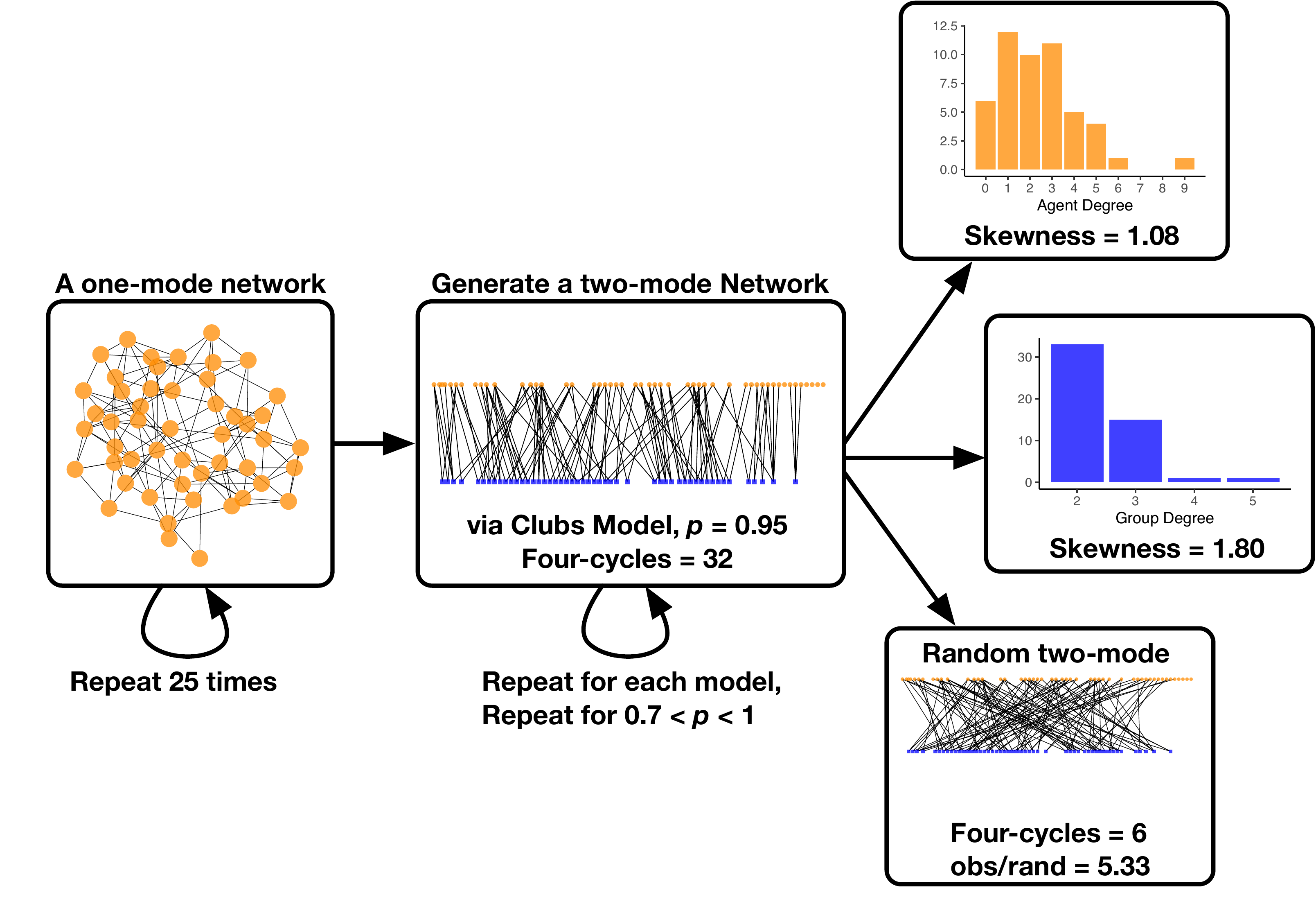}
    \caption{Evaluating a generated two-mode network.}
    \label{fig:design}
\end{figure}

Figure \ref{fig:design} illustrates how I examine whether the two-mode networks generated by these models have these empirically common characteristics. First, I generate a small-world network containing 50 nodes and 150 undirected edges. I begin with a small world network because it has properties that are observed in many real-world social networks (e.g., clustering, small mean distance). Second, I use the clubs model, with $p = 0.95$ to generate a two-mode network containing 50 groups. I choose to generate 50 groups because it keeps the experiment a manageable size, but large enough that each agent could be a member of a singleton group. In this generated two-mode network, most agents belong to just a few groups, and thus the agent degree distribution is positively skewed \citep[skewness = 1.08, using Fisher's moment coefficient of skewness;][]{joanes1998comparing}. Likewise, most groups have just a few members, and thus the group degree distribution is also positively skewed (skewness = 1.80). Finally, I use the curveball algorithm \citep{strona2014fast} to generate a random two-mode network with the same degree sequences, comparing the number of four-cycles in the generated and random networks. In this example, the generated network contains 5.33 times more four-cycles than a corresponding random network. Thus, in this example, the clubs model generated a two-mode network with all three expected properties.

\begin{figure}
    \centering
    \includegraphics[width=\textwidth]{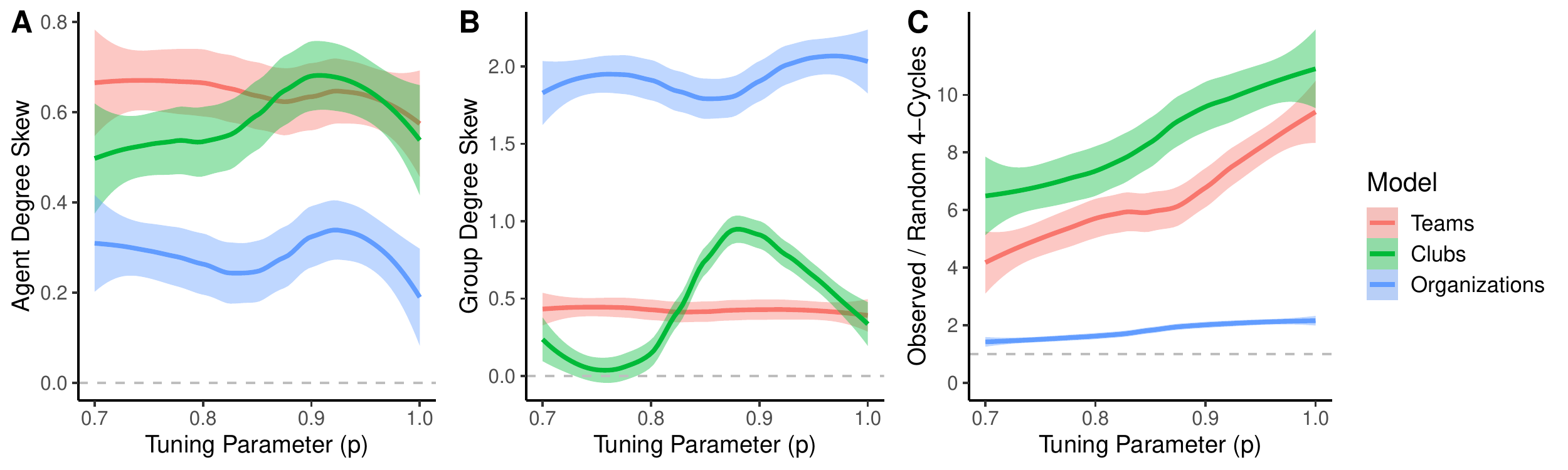}
    \caption{Experimental evaluation of generative models.}
    \label{fig:results}
\end{figure}

Figure \ref{fig:results} shows the results of repeating this evaluation process 25 times, for each generative model, and for each parameter $p$ between 0.7 and 1 in 0.025 intervals. Within each panel, the solid lines (red = teams model, green = clubs, blue = organizations) report averages over 25 replications, while the shaded bands indicate the 95\% confidence interval. Panel A illustrates that for all models and all values of $p$, the generated two-mode networks have a positively skewed agent degree distribution. Panel B illustrates that except for two mode networks generated using the clubs model with low values of $p$, all generated networks also have a positively skewed group degree distribution. Finally, panel C illustrates that for all models and all values of $p$, the generated two-mode networks have more four-cycles than a corresponding random network. Thus, collectively, this experiment demonstrates that under a broad set of circumstances, these models generate two-mode networks that have characteristics commonly observed in empirical two-mode social networks.

\subsection{Encoding one-mode networks}
\label{sec:encode}
The generative models all yield two-mode networks that have characteristics commonly observed in empirical two-mode social networks. However, the generative models should also yield two-mode networks that encode features of the particular one-mode networks from which they were generated. To evaluate this, I examine how well the original one-mode network can be recovered from the generated two-mode network.

\begin{figure}
    \centering
    \includegraphics[width=\textwidth]{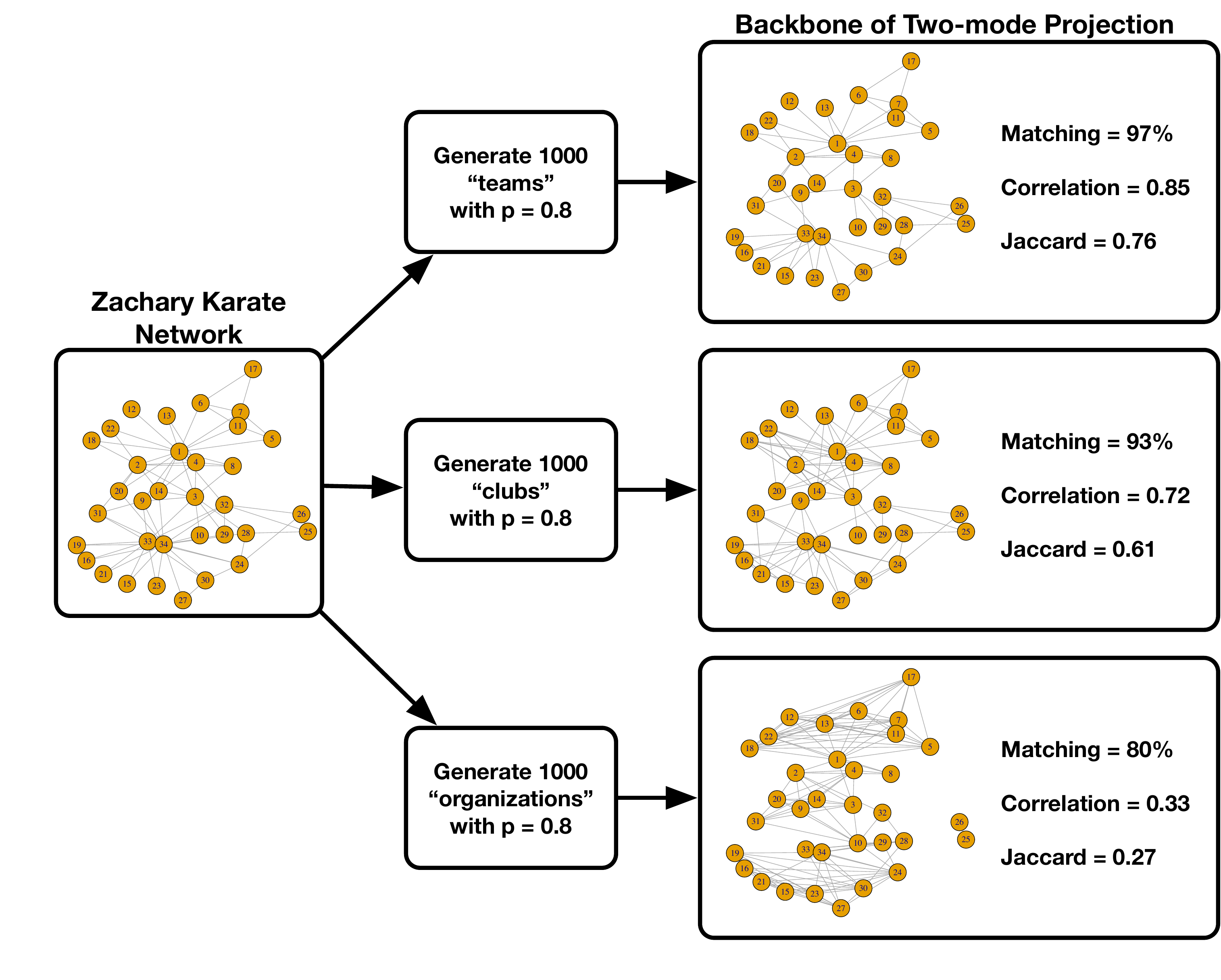}
    \caption{Recovering a one-mode network.}
    \label{fig:backbone}
\end{figure}

Using the \cite{zachary1977information} karate club network as the input, I use each model to generate a two-mode network of 1000 groups, with $p = 0.8$ (see Figure \ref{fig:backbone}). Setting $p = 0.8$ ensures that the generated two-mode networks contain a fair amount of noise and are not simply lists of cliques in the original network from which they are generated. Generating a large number of groups mirrors what a researcher might encounter when attempting to collect data in the field: an inability to directly observe the network of interest, but the ability to observe many instances of small events \citep[e.g.,][]{neal2022inferring}. For example, while it may be impossible to directly observe the karate club's social network, a researcher might be able to observe who participates in many small practice sessions and social events.

From each of the generated two-mode networks, I extract the backbone of its bipartite projection \citep{neal2014backbone}, then compute the similarity between this backbone and the original network. The simple matching coefficients (97\% -- 80\%) indicate that which dyads are (dis)connected in the backbone extracted from the generated two-mode networks closely matches which dyads are (dis)connected in the original one-mode network. More conservative similarity indices -- correlation (0.85 -- 033) and jaccard coefficient (0.76 -- 0.27) -- are expectedly lower, but are still positive and generally large. This analysis illustrates that the two-mode networks generated by these models are not simply random two-mode networks with empirically common features, but are two-mode networks that encode features of the specific one-mode networks from which they were generated.

\section{Discussion}
\label{sec:discussion}
Over a century ago, \cite{simmel} sketched the close association between individuals and groups. Building on these early ideas, \cite{breiger1974duality} demonstrated a method for deriving an interpersonal social network from individuals' group memberships, while \cite{feld1981focused} proposed focus theory to explain how social ties emerge from shared groups. Together, these methodological and theoretical contributions have facilitated research on how networks emerge from groups.

While prior work has provided the theoretical and methodological tools for understanding how groups lead to networks, less is known about the opposite process: \textit{how do networks lead to groups?} In this paper, building on ideas already present in focus theory and drawing on related theories of team \cite{guimera2005team}, club \cite{backstrom2006group,schaefer2022youth}, and organization \cite{mcpherson1983ecology} recruitment, I proposed three simple models for how a new group might emerge from an existing social network. In the \textit{teams} model, a new team is formed from incumbents of, and newcomers to, network cliques. In the \textit{clubs} model, a new club emerges as a members of a network clique attempt to recruit friends. Finally, in the \textit{organizations} model, a new organization recruit members from the interior and periphery of a socio-demographic niche.

These models can be viewed as two-mode network generative models, which are controlled by a tuning parameter $p$ that adjusts how closely the generated groups match the social network. Through a series of simulations, I demonstrated that these models generate two-mode networks that have characteristics commonly observed in empirical two-mode social networks: positively skewed agent degrees, positively skewed group degrees, and an over-representation of four-cycles. Additionally, using the \cite{zachary1977information} karate club network as an example, I illustrated that the generated two-mode networks are not simply random two-mode networks with empirically common features, but are two-mode networks that encode features of the one-mode network from which they were generated.

These models represent a theoretical contribution to the literature on networks and groups because they elaborate the missing second half of focus theory \citep{feld1981focused}. Specifically, while the influential focus theory hypothesized that groups (i.e., foci) lead to networks, and networks in turn lead to new groups, nearly all applications and extensions have focused on the first process, while neglecting the second process. To be sure, these models are simplified implementations of theories about group formation, and therefore are highly stylized. However, they provide a formalized starting point for further theoretical elaboration of focus theory, and of the co-evolution of networks and groups.

These models also represent a methodological contribution to the literature on network generative models. One-mode generative models -- for example, the Erd\H{o}s-R\'enyi \citep{erdosrenyi}, Watts-Strogatz \citep{watts1998collective}, and Barab\'asi-Albert \citep{barabasi1999emergence} models -- have played a critical role in understanding the properties of networks, and are frequently used as null models against which observed networks are evaluated. However, ``when it comes to bipartite networks [i.e., two-mode networks]...generative models are practically non-existent'' \citep[][p. 3]{vasques2020latent}. The generative models developed here, which yield two-mode networks with empirically common features and that encode features of one-mode networks, begin to fill that gap. Like existing generative models, they can be used to explore the properties of social two-mode networks, and can be used as null models against which observed two-mode networks are evaluated.

\subsection{Limitations and future directions}
These models and results are subject to some limitations, which highlight possible directions for future research. First, each model describes the emergence of a group solely from a network (i.e., network $\rightarrow$ group), and therefore does not allow individuals' participation in one group to influence their participation in future groups. More complex future models may allow groups to emerge not only as a function of the network, but also as a function of already existing groups (i.e., $^{\text{network}}_{\text{existing groups}} {^\rightarrow_\rightarrow}$ new group). Second, each model represents only a simplified implementation of a theory, and therefore does not attempt to incorporate all of the theory's mechanisms. For example, the organizations model is a significantly reduced form of organizational ecology, but provides a framework for future versions to incorporate additional elements such as niche carrying capacities \citep{popielarz2007niche} or competition \citep{shi2017member}. Third, the evidence that these models generate two-mode networks that contain features commonly observed in empirical two-mode networks is restricted to two types of features: skewed degree distributions and cycles. As future research identifies other common features of empirical two-mode networks, the simulations described in section \ref{sec:experiments} can be replicated to evaluate whether the generated two-mode networks also display these features. Finally, the conventional understanding of focus theory describes how a network can emerge from already-existing groups, while these models how a group can emerge from an already-existing network, and therefore neither offers a solution to the chicken-and-egg question: \textit{Which came first, groups or networks}? While this is likely a difficult question to answer, future work may integrate both processes to examine how groups and networks co-evolve as \cite{feld1981focused} initially hypothesized.

\subsection{Conclusions}
Theories \citep{feld1981focused} and methods \citep{breiger1974duality} have long acknowledged that individuals' group memberships can facilitate the formation of social ties. However, it is equally plausible that individuals' social ties can facilitate the formation of new groups. In this paper, I have sketched three models that describe how this might happen, and formalized them as two-mode generative models. These models have the potential to advance theories of how groups emerge from networks, as well as to provide methods for understanding and evaluating observed social two-mode networks. Moreover, as theoretically-informed but simple models, they also offer a starting point for the development of more complex and realistic models.

\section*{Data availability statement}
The code to replicate these analyses is available at 
\url{https://osf.io/eyua4/}.
%\url{https://osf.io/eyua4/?view_only=7b731b732cb540efb05947a535672e9f}.

\section*{Appendix. Model algorithms}
\begin{algorithm}
\caption{Teams model}\label{alg:team}
\textbf{Input: } $\mathbf{G} \gets$ Undirected, unweighted one-mode network\\
\textbf{Input: } $p \gets$ Tuning parameter (turnover probability) \\
\;
$incumbents \gets$ Members of random maximal clique $\in\mathbf{G}$\\
$S \gets |incumbents|$\\
$newcomers \gets$ Members of $\mathbf{G} \not\in incumbents$\\
$team \gets$ Random member $\in incumbents$\hfill {\footnotesize $\triangleright$ First position filled}\\
$incumbents \gets incumbents - team$\hfill {\footnotesize by an incumbent}\\
\For(\hfill {\footnotesize $\triangleright$ Remaining positions filled by:}){\upshape{$position$ from 2 to $S$}}
        {\eIf(\hfill {\footnotesize (A) Incumbents, or}){$rand() \leq p$}
            {$i \gets$ Random member $\in incumbents$\\
            $team \gets team + i$\\
            $incumbents \gets incumbents - i$}
            (\hfill {\footnotesize (B) Newcomers}){$i \gets$ Random member $\in newcomers$\\
            $team \gets team + i$\\
            $newcomers \gets newcomers - i$}}
\;
\textbf{Output: } $team$, list of team members
\end{algorithm}

\begin{algorithm}
\caption{Clubs model}\label{alg:club}
\textbf{Input: } $\mathbf{G} \gets$ Undirected, unweighted one-mode network\\
\textbf{Input: } $p \gets$ Tuning parameter (minimum club density) \\
\;
$club \gets$ Members of random maximal clique $\in \mathbf{G}$\hfill {\footnotesize $\triangleright$ Group begins with random clique}\\
$declined \gets$ \{\}\\
$candidates \gets$ Neighbors of $club \in \mathbf{G}$\hfill {\footnotesize $\triangleright$ Members recruit friends}\\

\While{\upshape{$|candidates| \neq 0$}}
        {$recruit \gets$ Random member $\in candidates$\\
        $\mathbf{G}' \gets$ Subgraph of $\mathbf{G}$ containing $members$ and $recruit$\\
        {\eIf(\hfill {\footnotesize $\triangleright$ A recruit either:}){density($\mathbf{G}'$) $\geq p$}
            {$club \gets club + recruit$\hfill {\footnotesize (A) Joins, or}}
            {$declined \gets declined + recruit$\hfill {\footnotesize (B) Declines}}}
        $candidates \gets$ Neighbors of $club \in \mathbf{G} - declined$}
\;
\textbf{Output: } $club$, list of club members
\end{algorithm}

\begin{algorithm}
\caption{Organizations model}\label{alg:organization}
\textbf{Input: } $\mathbf{G} \gets$ Undirected, unweighted one-mode network\\
\textbf{Input: } $p \gets$ Tuning parameter (probability of recruitment success)\\
\textbf{Input: } $d \gets$ Dimensionality of Blau Space \hfill {\footnotesize $\triangleright$ Usually $d=2$}\\
\;
$\mathbf{G}' \gets$ Geodesic distances in $\mathbf{G}$\hfill {\footnotesize $\triangleright$ Obtain Blau space}\\
$\mathbf{D} \gets$ Euclidean distances in $d$-dimensional embedding of $\mathbf{G}'$\\
$R \gets$ Random draw from a positive-skew distribution \hfill {\footnotesize $\triangleright$ Pick niche radius}\\
$R \gets (R \times (max(D) - min(D))) + min(D)$ \\
$center \gets $ Random agent $\in \mathbf{G}$ \hfill {\footnotesize $\triangleright$ Pick niche center}\\
$inside \gets i \in \mathbf{G}$ where $D_{i,center} \leq R$ \hfill {\footnotesize $\triangleright$ Agents inside niche} \\
$outside \gets i \in \mathbf{G}$ where $D_{i,center} > R$ \hfill {\footnotesize $\triangleright$ Agents outside niche} \\
\;
\For(\hfill {\footnotesize $\triangleright$ Recruit from inside niche}){each $i \in inside$}
  {\If{$rand() \leq p$}{$org \gets org + i$}}
\;
\While(\hfill {\footnotesize $\triangleright$ If spaces remain,}){$|org| < |inside|$}
    {$i \gets$ agent with $min(D_{i,center}) \in outside$ \hfill {\footnotesize recruit from outside niche}\\
    \textbf{if} $rand() \leq p$ \textbf{then} $org \gets org + i$\\
    $outside \gets outside - i$}
\;
\textbf{Output: } $org$, list of organization members
\end{algorithm}

\end{document}